\documentclass[prd,amstext,amsmath,amssymb,superscriptaddress,floatfix,nofootinbib,notitlepage]{revtex4-1}
\usepackage[hyperref]{xcolor}
\usepackage{graphics}
\usepackage{epsfig}
\usepackage{amsmath}
\usepackage{amssymb}
\usepackage{amsthm}
\usepackage[mathscr]{eucal}
\usepackage{graphicx}
\usepackage{subfigure}
\usepackage{bbm}
\usepackage{ulem}
\usepackage{slashed}
\usepackage{hyperref}
\hypersetup{
  colorlinks, breaklinks,
  linkcolor=blue,
  citecolor=blue,
  linktoc=all
}

\newcommand{\Eq}[1]{{Eq.~({\ref{#1}})}}
\newcommand{\Sect}[1]{{Sec.~{\ref{#1}}}}
\newcommand{\Eqs}[1]{{Eqs.~({\ref{#1}})}}
\newcommand{\Fig}[1]{{Fig.~{\ref{#1}}}}

\newcommand{\bea}{\begin{eqnarray}}
\newcommand{\eea}{\end{eqnarray}}
\newcommand{\beq}{\begin{equation}}
\newcommand{\eeq}{\end{equation}}
\newcommand{\beas}{\begin{eqnarray*}}
\newcommand{\eeas}{\end{eqnarray*}}

\newcommand{\equalover}[1]{{\boldmath{\overset{#1}{\,=}\,}}}
\newcommand{\lraunder}[1]{{\boldmath{\underset{#1}{\;\longrightarrow}\;}}}

\def\p{{\bf p}}

\def\x{{\bf x}}
\def\ms{\ensuremath{m_\sigma}}
\def\msc{m_{\sigma,c}}
\def\msp{m_{\sigma,p}}
\def\msBC{m_{\sigma,\mathrm{BC}}}
\def\msRP{m_{\sigma,\mathrm{RP}}}

\def\avgs{{\sigma}}

\def\Mv{M_{\text{v}}}
\def\svac{\sigma_{\text{{v}}}}

\newcommand{\Lonez}{{L_1\big|_{M=0}}}
\newcommand{\Lonem}{{L_1}}
\newcommand{\Lonemv}{{L_1}}
\newcommand{\Ltwoz}{{L_2(0)\big|_{M=0}}}
\newcommand{\Ltwom}{{L_2(0)}}
\newcommand{\Ltwomv}{{L_2(0)}}

\newcommand{\Ltwomsmv}{{L_2(\ms^2)}}
\newcommand{\Ltwomspmv}{{L_2(\msp^2)}}

\graphicspath{
{./}
{./figures/}
}

\begin{document}

\title{
Consistent parameter fixing in the quark-meson model  with vacuum fluctuations
  }

\author{Stefano~Carignano}
\affiliation{ INFN,  Laboratori  Nazionali  del  Gran Sasso, Assergi (AQ), Italy}

\author{Michael~Buballa}
\affiliation{Theoriezentrum, Institut f\"ur Kernphysik, Technische Universit\"at Darmstadt, Germany}

\author{Wael~Elkamhawy}
\affiliation{Theoriezentrum, Institut f\"ur Kernphysik, Technische Universit\"at Darmstadt, Germany}

\begin{abstract}

We revisit the renormalization prescription for the quark-meson model in an extended mean-field approximation, where vacuum quark fluctuations are included. 
At a given cutoff scale the model parameters are fixed by fitting vacuum quantities, 
typically including the sigma-meson mass $\ms$ and the pion decay constant $f_\pi$.
In most publications the latter is identified with the expectation value of the sigma field, while for $\ms$ the curvature
mass is taken. When quark loops are included, this prescription is however inconsistent, and the correct identification involves 
the renormalized pion decay constant and the sigma pole mass. In the present article we investigate the influence of the 
parameter-fixing scheme on the phase structure of the model at finite temperature and chemical potential.
Despite large differences between the model parameters in the two schemes, we find that in homogeneous matter the effect on the phase diagram is relatively small.
For inhomogeneous phases, on the other hand, the choice of the proper renormalization prescription is crucial.  	 
In particular, we show that if renormalization effects on the pion decay constant are not considered, the model does not even present a well-defined renormalized limit when the cutoff is sent to infinity. 
\end{abstract}

\maketitle

\section{Introduction}

The study of strong-interaction matter at finite density is a fascinating topic which attracts a lot of interest, both from the theory and the experimental side.
While  weak-coupling
 calculations reveal
 that the ground state of quantum chromodynamics (QCD) at asymptotically high baryonic chemical potentials and low temperatures is a color superconductor, the region of intermediate densities beyond nuclear matter saturation may exhibit a rich phase 
 structure, see e.g.~\cite{Fukushima:2010bq} for a review.

From a theoretical point of view, the investigation of this region is extremely challenging: due to the sign problem, 
ab-initio lattice simulations are unavailable at nonzero baryonic densities, so that 
phenomenological models are currently the main tool to access this region.  
A popular choice for this kind of study is the quark-meson (QM) model \cite{Scavenius:2000qd, Schaefer:2006ds}, whose basic building blocks are 
quarks which interact with mesons. Compared to the formally similar Nambu--Jona-Lasinio (NJL) model, the QM model has the advantage of being renormalizable, a property which renders it an excellent tool for investigating the role of fluctuations in a systematic way within the framework of the functional renormalization group, e.g.~\cite{Schaefer:2004en,Schaefer:2006ds,Skokov:2010wb,Herbst:2013ail,Tripolt:2013zfa}.

Until a few years ago it was believed that, in the mean-field approximation, quark vacuum fluctuations could be simply left out, expecting that a redefinition of the meson potential parameters would be sufficient to take their effects into account. Such a ``no-sea'' or ``standard mean-field approximation'' (sMFA) leads however to inconsistent predictions, such as the persistence of a first-order chiral phase transition in the chiral limit at zero densities, in conflict with
general expectations~\cite{Pisarski:1983ms,Halasz:1998}.  
This artifact
disappears once vacuum quark fluctuations are properly taken into account \cite{Skokov:2010sf,Schaefer:2011ex,Gupta:2011ez}, suggesting that an explicit treatment of the Dirac sea contributions is crucial.

 More recently, this so-called extended mean-field approximation (eMFA) where mesonic fluctuations are still neglected
  has also been applied to study the effects of quark vacuum fluctuations on inhomogeneous 
 chiral symmetry breaking phases~\cite{Carignano:2014jla}. 
 These crystalline phases, which are  characterized by the formation of a spatially modulated quark-antiquark condensate, 
 are expected to form in cold and dense quark matter (for a recent review, see \cite{Buballa:2014tba}). 
Mean-field NJL-model studies suggest that they completely cover the first-order chiral transition~\cite{Nickel:2009ke}, so that
 the critical point (CP) is replaced by a Lifshitz point (LP), denoting the tip of the inhomogeneous island. 
 For the QM model it was found that  including the Dirac sea reduces the size of the inhomogeneous phase, but in general
 does not destroy it completely~\cite{Carignano:2014jla}. 
In particular, it was shown that the LP is at the same place as the CP of the homogeneous analysis
if in vacuum the sigma-meson mass $\ms$ is twice the constituent quark mass, as it is always the case  in the NJL model.
In the QM model, however,  $\ms$ can be chosen freely, and the inhomogeneous phase turned out to be very sensitive
to this choice.

In the eMFA the diverging contribution stemming from vacuum quark fluctuations 
 are reabsorbed by a proper redefinition of the model parameters through a fit to vacuum observables, leading to UV-finite results. 
As we shall see, however, this renormalization procedure must be performed with great care, 
as different prescriptions for the identification of the physical quantities can be employed.

In its simplest incarnation, the two-flavor QM model in the chiral limit has three free parameters, which are typically fitted in vacuum to give reasonable values of the pion decay constant $f_\pi$, the sigma meson mass $\ms$ and the constituent quark mass $\Mv$.  
Until relatively recently, the standard procedure has been to fit the sigma meson mass
to the so-called ``curvature'' (or screening) mass, associated with 
the curvature of the QM thermodynamic potential at its minimum, while the pion decay constant is identified with the vacuum expectation value of the sigma mean-field \cite{Skokov:2010sf,Chatterjee:2011jd,Gupta:2011ez,Herbst:2013ail,Andersen:2012bq,Weyrich:2015hha}.
 In principle, however, the physical mass of the sigma meson is given by the pole of its propagator (as recently stressed in \cite{Strodthoff:2011tz,Kamikado:2012bt}), while the pion decay constant is related to the residue of the pion propagator at its pole.
In light of these considerations, it was suggested in \cite{Carignano:2014jla} to consider these ``pole'' quantities instead of the traditionally employed ones for the parameter fixing in the model.
While in absence of Dirac sea contributions the two prescriptions become trivially equivalent, when quark loops are taken into account  the sigma pole and screening masses start to differ, and the pion decay constant has to be renormalized as well.

The main objective of this work is to investigate the differences between the extended mean-field results obtained in the QM model within these two different prescriptions, 
with a particular emphasis on their influence on inhomogeneous chiral symmetry breaking phases.
While, as we shall discuss, in the QM model homogeneous phases turn out to be relatively insensitive to the specific parameter-fixing scheme employed, for inhomogeneous phases this choice becomes crucial. 

The remainder of this article is organized as follows. In Sect.~\ref{sec:qmmodel} we introduce the model 
and give a brief summary of the basic formalism employed in \cite{Carignano:2014jla} to study the phase diagram.
After that, in Sect.~\ref{sec:schemes}, we define two different parameter-fixing schemes and then compare the resulting phase 
diagrams in Sect.~\ref{sec:comparison}.
In Sect.~\ref{sec:renormGL}  we perform a Ginzburg-Landau analysis for the CP and the LP, focusing on the renormalized limit.
Two further parameter-fixing schemes are briefly discussed in Sect.~\ref{sec:otherschemes}, before we draw our conclusions
in Sect.~\ref{sec:conclusions}.

\section{Extended mean-field approach in the quark-meson model}
\label{sec:qmmodel}

Before introducing the different parameter fixing prescriptions, we set the stage by reviewing 
the basic formalism needed for our discussion.
 A more detailed derivation can be found in \cite{Carignano:2014jla}.

The quark-meson model Lagrangian density is given by
\cite{Scavenius:2000qd,Schaefer:2006ds}
\beq
\mathcal{L}_\text{QM}
=
\bar{\psi}
\left(
i\gamma^\mu \partial_\mu
-
g(\sigma+i\gamma_5 \vec\tau \cdot \vec\pi)
\right)
\psi
+
\mathcal{L}_{\text{mes}}
\ ,
\eeq
where $\psi$ is a $4N_f N_c$-dimensional quark spinor with $N_f = 2$
flavor and $N_c = 3$ color degrees of freedom, $\sigma$ is the scalar
field of the sigma meson and $\vec \pi$ the pseudo-scalar fields of the
pion triplet. 
 The purely mesonic term $\mathcal{L}_{\text{mes}}$ contains a kinetic and a potential term,
\beq
\mathcal{L}_{\text{mes}} = \frac{1}{2} \left(
\partial_\mu\sigma \partial^\mu\sigma 
+
\partial_\mu\vec\pi \partial^\mu\vec\pi
\right) 
- U(\sigma,\vec\pi ) 
\label{eq:Lmeson}
\eeq
with
\beq
U(\sigma,\vec\pi ) = 
\frac{\lambda}{4}
\left(
\sigma^2
+
\vec\pi^2
- 
v^2
\right)^{2} - c \sigma 
\,.
\eeq

In the following we will work in the chiral limit by considering $c=0$. 
The meson potential $U(\sigma,\vec\pi)$ has then an exact $O(4)$-symmetry, which is isomorphic to the $SU(2)_L \times SU(2)_R$ chiral symmetry, and the Lagrangian is characterized by 
 three model parameters, $g$, $\lambda$
and $v^2$.

The thermodynamic properties of the model are encoded in the grand
potential $\Omega$. 
  In mean-field
approximation we treat the meson fields $\sigma$ and $\pi^{a}$ as
classical and replace them by their expectation
values~\cite{Scavenius:2000qd, Schaefer:2006ds}, 
which we assume to be static.  Using standard techniques,
  the thermodynamic potential can then be written as the sum of a 
   quark loop contribution $\Omega_{q}$ and a pure mesonic term $\Omega_\text{mes}$.
In particular,  after combining the mesonic mean fields with the Yukawa coupling $g$
(as detailed below for two specific examples)
one finds that the quark contribution has the same form as in the NJL model,
and does not depend explicitly on the QM model parameters~\cite{Buballa:2014tba}. 
   
In our discussion we will consider both the standard homogeneous case where $\sigma$ is taken to be spatially constant and $\pi^a=0$, 
as well as the case where the meson mean-fields are inhomogeneous.
In the former it is convenient 
to define $\Delta = g\sigma$, which can be interpreted as a constituent quark mass. In this case, the meson contribution to the thermodynamic 
potential is
\beq
\Omega_{\text {mes}}^{\text{hom}}(\sigma, \vec\pi) =
 U\left(\frac{\Delta}{g}, \vec{0}\right) 
 =
 \frac{\lambda}{4}\left[\left(\frac{\Delta}{g}\right)^2 - v^2\right]^2 
\equiv
U(\Delta)
\,.
\label{eq:UDelta}
\eeq

Since at $T=\mu=0$ the mean-fields are expected to be homogeneous, we can employ this ansatz to calculate the vacuum thermodynamic potential, which is given by
\beq
\label{eq:OmegaQMVacpisigma}
\Omega^\text{vac}(\sigma, \vec\pi) = 
- 2 N_f N_c
\int\!\! \frac{d^3p}{(2\pi)^3}\, E_\p 
+ U(\Delta) \,,
\eeq
where $E_\p = \sqrt{\p^2 + \Delta^2}$ and the first term corresponds to the contribution due to vacuum fluctuations of quarks, which plays a crucial role in our discussion.

This quark loop integral is quartically divergent and needs to be regularized. Here we employ a Pauli-Villars (PV)-inspired scheme, 
which is considered appropriate for dealing both with homogeneous and inhomogeneous solutions and amounts to the replacement \cite{Nickel:2009wj}
\beq
\label{eq:Epv}
E_\p \rightarrow \sum_j c_j \sqrt{ E_\p^2 + j \Lambda^2 } \,, \qquad c_j = \{1, -3, 3, -1\}  \,,
\eeq
 where $\Lambda$ is the PV regulator.
In particular, the sMFA results are recovered for $\Lambda =0$, while for nonzero values of $\Lambda$ effects of the
 Dirac sea are included.
It is important to recall at this point that the QM model is renormalizable. Therefore, although the thermodynamic potential always depends on the chosen value of $\Lambda$, the model results should eventually become independent of the regulator when it is sufficiently large. 
 This aspect will be investigated when discussing the different parameter fixing prescriptions.

At the minimum, where the condition  $\partial\Omega_\text{vac}/{\partial\sigma} = 0$ must hold, 
the nontrivial solution for the sigma field, which we will denote as $\svac$, satisfies the gap equation 
\beq
\label{eq:gapL1}
 \lambda \left(\frac{\Mv^2}{g^2} - v^2\right)
=
g^2\,\Lonemv\,,
\eeq
where we also defined the vacuum constituent quark mass as
$\Mv \equiv g\svac$. $L_1$ is a quadratically divergent loop integral which is
regularized consistently with \Eq{eq:Epv}, see Appendix \ref{app:Lint}.

Our ansatz for spatially inhomogeneous matter will be 
 a one-dimensional chiral density wave (CDW),
 which is given by \cite{Dautry:1979bk,Kutschera:1989yz,Nakano:2004cd}
(see also \cite{Buballa:2014tba} for a more detailed discussion)
\beq
\avgs(z) = \frac{\Delta}{g} \cos(2 i q z)\,,\quad  
\pi^3 (z) = \frac{\Delta}{g} \sin(2 i q z)  
\,, \qquad \pi^1 = \pi^2 = 0 \,.
\label{eq:cdwdef}
 \eeq

With this type of inhomogeneous order parameter, the evaluation of the quark loop contributions to the thermodynamic potential becomes more involved but is still feasible. 
Compared to homogeneous matter, the mesonic contribution now contains an additional kinetic term,
\beq
\Omega_\text{mes}^{\text{CDW}}(\sigma, \vec\pi)  = 
 \frac{1}{2} \left(\frac{2q \Delta}{g}\right)^2 + U(\Delta)\,,
 \label{eq:mesonpotCDW}
\eeq
and we recover the homogeneous limit for
 $q = 0$ (cf. also \Eq{eq:cdwdef}).
In spite of its simplicity, the CDW constitutes an excellent prototype to gauge the sensitivity of inhomogeneous phases with respect to the different parametrizations considered 
within the QM model, as previous studies have shown that the resulting phase structure for different types of spatial modulations is similar. In particular, it was shown that the position of the Lifshitz point 
and the second-order phase boundary to the restored phase are general results
independent of the particular spatial dependence chosen for the mean fields \cite{Nickel:2009ke,Nickel:2009wj}.  This can be seen within the context of a Ginzburg-Landau (GL) analysis, which provides a systematic framework to investigate the thermodynamic potential in vicinity of a second-order phase transition \cite{Nickel:2009ke,Nickel:2009wj, Abuki:2011pf}. %

For this, we expand the thermodynamic potential in terms of a complex constituent-quark mass function, which in our case is
 given by $M(\x)=g(\sigma(\x) + i\pi^3(\x))$, and obtain
\beq
       \Omega
       (T,\mu;M(\x) ) =
        \Omega
        (T,\mu;0)
       + \frac{1}{V}\int d^3x\, \left\{\frac{1}{2}\gamma_2 |M(\x)|^2
       + \frac{1}{4}\gamma_{4,a} |M(\x)|^4 + \frac{1}{4}\gamma_{4,b} |\nabla M(\x)|^2
       + \dots
       \right\}\,.
\label{eq:OmegaGL}
\eeq
Within this setup, the location of the CP is  determined by
the condition that the coefficients of the quadratic and quartic terms
vanish, ie. $\gamma_2 \big|_\text{CP} = \big. \gamma_{4,a}
\big|_\text{CP} = 0$,  while at the LP the quadratic and the gradient terms are zero, $\big. \gamma_2 \big|_\text{LP} = \big. \gamma_{4,b}
\big|_\text{LP} = 0$ \cite{Nickel:2009ke}.

Given the structure of the thermodynamic potential, the GL coefficients $\gamma_i $ can be split into a pure mesonic contribution $\alpha_i$  as well as a quark-loop one $\beta_i$, i.e., $\gamma_i = \alpha_i + \beta_i$ \cite{Carignano:2014jla}.
The mesonic coefficients $\alpha_i$ are easily obtained from \Eq{eq:Lmeson}, 
yielding~\cite{Nickel:2009wj}
\beq
       \alpha_2 = -\frac{\lambda v^2}{g^2}\,, \quad
       \alpha_{4,a} = \frac{\lambda}{g^4}\,, \quad
       \alpha_{4,b} = \frac{2}{g^2}\,, 
       \label{eq:glalphas}
\eeq
while the quark-loop terms $\beta_i$ have the same structure as in 
the NJL model~\cite{Nickel:2009wj, Nickel:2009ke}.
 They can be written as
\bea
\beta_2 &=& \beta_2^\text{vac} + \beta_2^\text{med}\,,
\\
\beta_{4,a} = \beta_{4,b} &=& \beta_4^\text{vac} + \beta_4^\text{med}\,,
\eea
where the medium contributions contain Fermi distribution functions and are always UV-finite, while the vacuum contributions are given by
\bea
        \beta_2^\text{vac} &=& \big.-\Lonez \,,
\\[1ex]
        \beta_4^\text{vac} &=& \big.-\Ltwoz \,, 
        \label{eq:glbetasvac}
\eea
with the regularized integrals $L_1$ and $L_2$, as described in the appendix.
As for the full thermodynamic potential, we note that here only the mesonic coefficients $\alpha_i$ depend explicitly on the QM model parameters, while the entire $T$ and $\mu$ 
dependence of the coefficients lies in the quark contributions $\beta_i^{\text{med}}$. 
Since in the NJL model there are no mesonic terms\footnote{Instead there is an additional condensate term in the NJL model, which however only contributes to $\gamma_2$. }
 and $\beta_{4,a}=\beta_{4,b}$, there one finds that the positions of the CP and the LP coincide \cite{Nickel:2009ke}. 
In \cite{Carignano:2014jla}
it was found that this is also true in the QM model if the vacuum value of the sigma meson mass is twice as big as the
constituent quark mass.
In Sect.~\ref{sec:comparison}  we will discuss how this result depends
on the parameter fixing  schemes defined in the next section.

\section{Parameter fixing prescriptions}
\label{sec:schemes}

The three parameters of the QM model, $g^2$, $\lambda$ and $v^2$,
are determined by fitting three ``observables'', the pion decay constant $f_\pi$, the constituent quark mass $\Mv$ and 
the sigma mass $m_\sigma$ in vacuum.
Admittedly, the sigma mass and in particular the constituent quark mass do not correspond to good observables in reality, 
but this is unimportant for our discussion. 
Here we assume that empirical values for these quantities exist, which then serve as input for the fitting procedure.\footnote{For the sake of clarity, we specify that  in the following $f_\pi$ and $m_\sigma$ always refer to the input numbers, while we will use different names for the model
expressions which are identified with these quantities.}

The most commonly employed prescription in the literature associates the pion decay constant $f_\pi$  
with  
the vacuum expectation value of the sigma field $\svac$,
while $\ms$ is identified with the sigma curvature mass $\msc$.
The latter
is defined through the equation 
\beq
\label{eq:mscreen}
      m_{\sigma,c}^2 
      \equiv     
       \left. \frac{\partial^2\Omega_\text{vac}}{\partial\sigma^2}
       \right|_{\sigma = \svac, \vec\pi=0}
       = 
       -2g^2 \Mv^2 \Ltwomv + 2\lambda \frac{\Mv^2}{g^2}\,,
\eeq 
where the second equality is obtained by taking the second derivative of \Eq{eq:OmegaQMVacpisigma} and 
using the gap equation \Eq{eq:gapL1}. 
 This scheme, which we will refer to as "BC" (as in Bare $f_\pi$ and Curvature mass), is thus defined by  
\beq
\text{BC:} \qquad
       \svac  \equiv \frac{\Mv}{g} \equalover{!} f_\pi\,,
       \qquad
       m_{\sigma,c}^2 \equalover{!} m_\sigma^2 \,.
\eeq
From the first equation one trivially obtains an expression for $g$
\beq
       g^2 = \frac{\Mv^2}{f_\pi^2} ,
\label{eq:g2BC}
\eeq
which can then be inserted into \Eq{eq:mscreen}, 
so that we get 
\beq
       \lambda = \frac{m_\sigma^2 f_\pi^2 + 2\Mv^4 \Ltwomv}{2 f_\pi^4}.
\label{eq:lambdaBC}
\eeq
Finally, we use the gap equation (\ref{eq:gapL1}) to solve for $v^2$ as function of the other two parameters:
\beq
       v^2 = \frac{\Mv^2}{g^2} - \frac{g^2 \Lonemv}{\lambda}  \,.
       \label{eq:v2}
\eeq

While the BC prescription is consistent in the sMFA, this is no longer the case
once vacuum quark fluctuations are present.
Since the latter correspond to loop corrections to the gap equation, 
one has to consider such corrections to the mesons as well.
The resulting dressed meson propagators then take the form
\beq
       D_j(q^2) = \frac{1}{q^2 - m_{j,t}^2 +  g^2 \Pi_j(q^2) +i\epsilon}\ ,
\label{eq:DM}
\eeq
where $j = \sigma, \pi$ denotes the meson channel,
$m_{j,t}$ is the corresponding tree-level mass, 
and $\Pi_j(q^2)$ describes the $q\bar q$ polarization function in this channel.
In the vicinity of the pole this can be written as
\beq
       D_j(q^2) = \frac{Z_j}{q^2 - m_{j,p}^2 +i\epsilon} \; + \; \text{regular terms}\,,
\eeq
with the pole mass $m_{j,p}$ implicitly defined by $D_j^{-1}(m_{j,p}) = 0$
and the wave-function renormalization constant 
$Z_j^{-1} = 1+ g^2 (d\Pi_j/dq^2)|_{q^2=m_{j,p}^2}$.

In contrast to the curvature mass, the pole mass corresponds to an observable quantity 
and should therefore be used in the renormalization procedure.
For the sigma meson one finds~\cite{Carignano:2014jla}
\beq
       D_\sigma^{-1}(m_{\sigma,p}^2) = 
       m_{\sigma,p}^2 
       -\frac{1}{2} g^2 (m_{\sigma,p}^2 -4\Mv^2) \Ltwomspmv 
       -2\lambda \frac{\Mv^2}{g^2}       
       = 0\,,
\label{eq:msigmapole}
\eeq
from which $\msp$ can be determined.\footnote{Special care must be taken for  $\msp > 2\Mv$ since $L_2$ gets a nonvanishing imaginary part due to the
open $\sigma \rightarrow q\bar{q}$ decay channel in the model. In this case we define 
$\msp$ via the real part of $L_2$, but our main conclusions will not depend on this definition
. } 
In the sMFA we have $L_2= 0$, and hence $\msp = \msc$, cf.~\Eq{eq:mscreen}.  
In general, however, pole and curvature masses are different.

In the pion channel, on the other hand, one finds $m_{\pi,p} = m_{\pi,c}= 0$,
i.e., both prescriptions are consistent with the Goldstone theorem. 
Nevertheless, as a consequence of the loop corrections, one gets $Z_\pi \neq 1$, 
which leads to a renormalization of the pion decay constant.
One obtains\footnote{
Most easily, this can be motivated by the Goldberger-Treiman relation 
$\Mv =  g_\text{ren} f_{\pi,\text{ren}}$ where $g_\text{ren} = g \sqrt{Z_\pi}$ is the renormalized quark-pion 
coupling~\cite{Carignano:2014jla}.
More rigorously, the pion decay constant is obtained from the pion-to-vacuum matrix element after 
coupling the model to an axial gauge field. From this it follows that the decay constant is directly
related to the pion wave function and must be renormalized accordingly. 
} \cite{Carignano:2014jla}
\beq
       f_{\pi,\text{ren}}^2 =  \frac{\svac^2}{Z_\pi}= \frac{\Mv^2}{g^2} \left( 1  - \frac{1}{2} g^2\Ltwomv\right) \,,
\label{eq:fpiren}
\eeq
which can then be identified with the physical value $f_\pi^2$.

The prescription proposed in \cite{Carignano:2014jla}, which employs these quantities and which we will refer to as RP scheme (as in Renormalized $f_\pi$ and Pole mass) 
is therefore defined as 
\beq
\text{RP:} \qquad 
       f_{\pi,ren}^2 \equalover{!} f_\pi^2\,,
       \qquad
       m_{\sigma,p}^2 \equalover{!} m_\sigma^2 \,.
\eeq
Using \Eqs{eq:fpiren} and (\ref{eq:msigmapole}), this yields
\beq
       g^2 = \frac{\Mv^2}{f_\pi^2 + \frac{1}{2}\Mv^2 \Ltwomv}  
\label{eq:g2RP}
\eeq
and
\beq
      \lambda
       =
       2g^2 \frac{m_{\sigma}^2}{4\Mv^2} 
       \left[ 1
       - \frac{1}{2}g^2
       \left(1 - \frac{4\Mv^2}{m_{\sigma}^2} \right) \Ltwomsmv  
       \right]\,,
\label{eq:lambdaRP}
\eeq
while $v^2$ is again obtained from the gap equation (\ref{eq:gapL1}) (cf. \Eq{eq:v2}).

\section{Comparison of the schemes}
\label{sec:comparison}

\begin{figure}

\hspace{.1cm}\includegraphics[width=.3\textwidth]{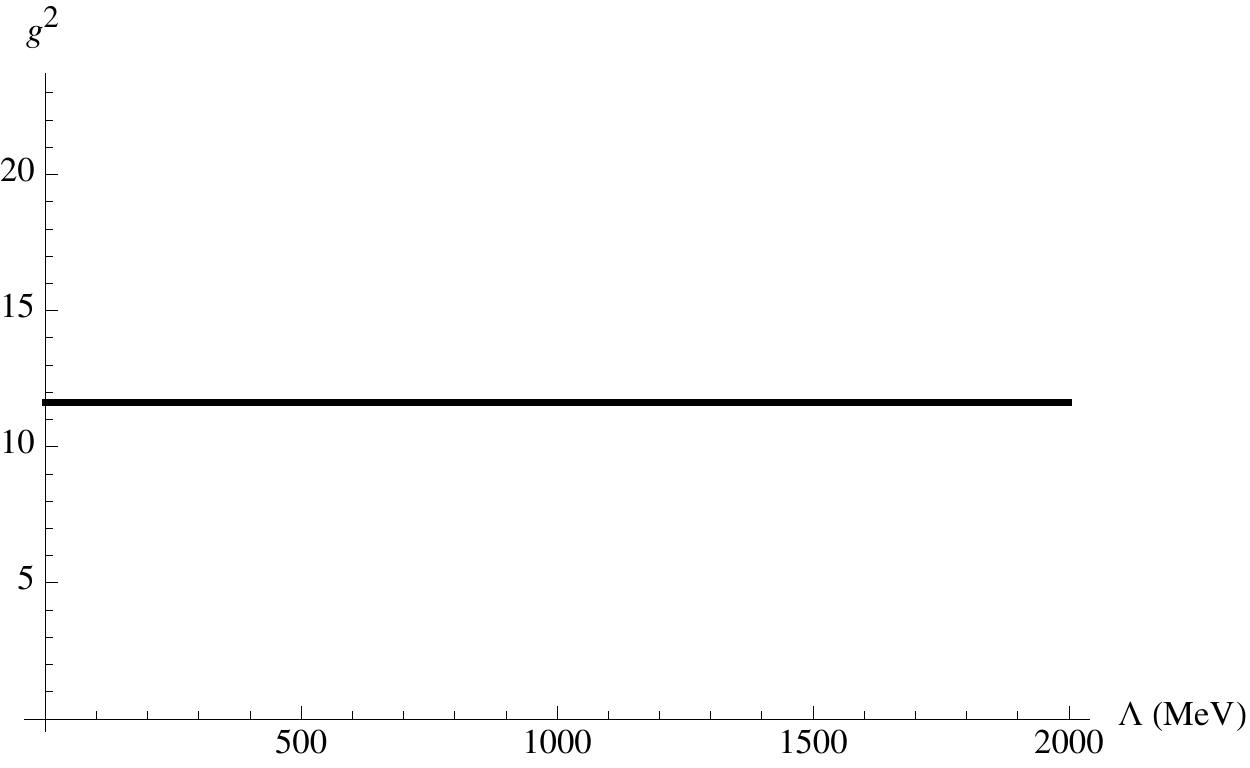}
\hspace{-0.03cm}\includegraphics[width=.3\textwidth]{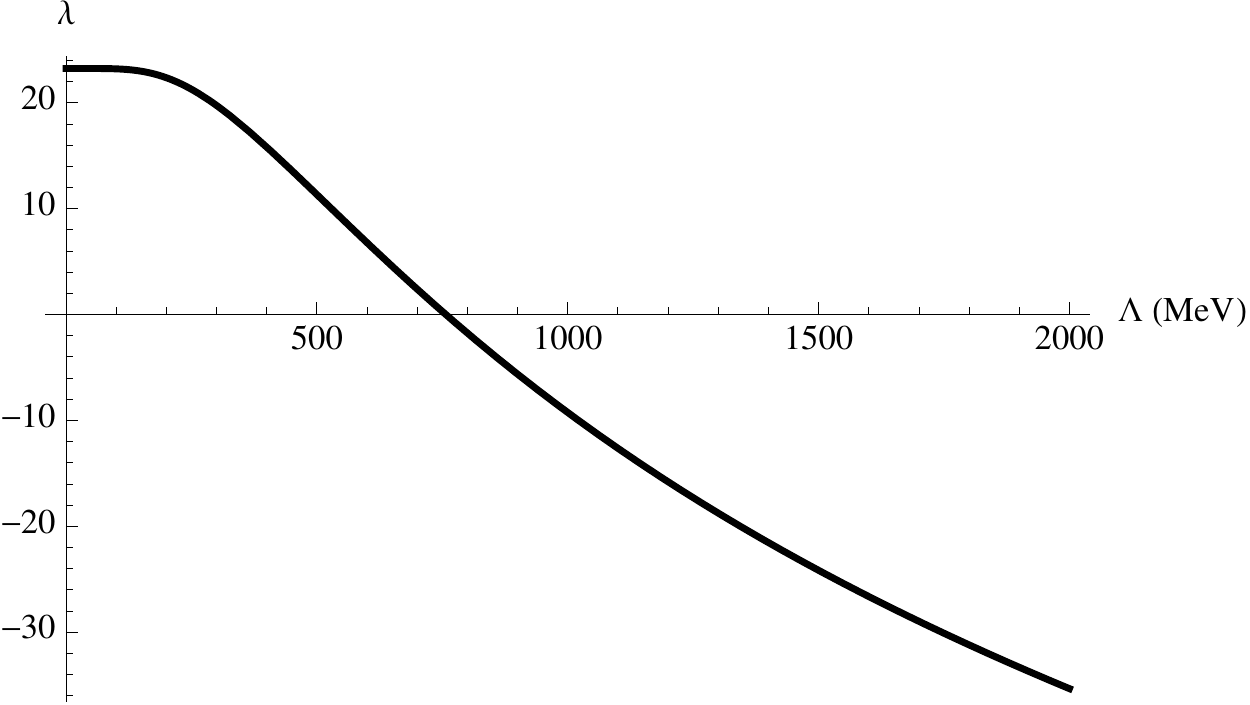}
\hspace{-0.08cm}\includegraphics[width=.3\textwidth]{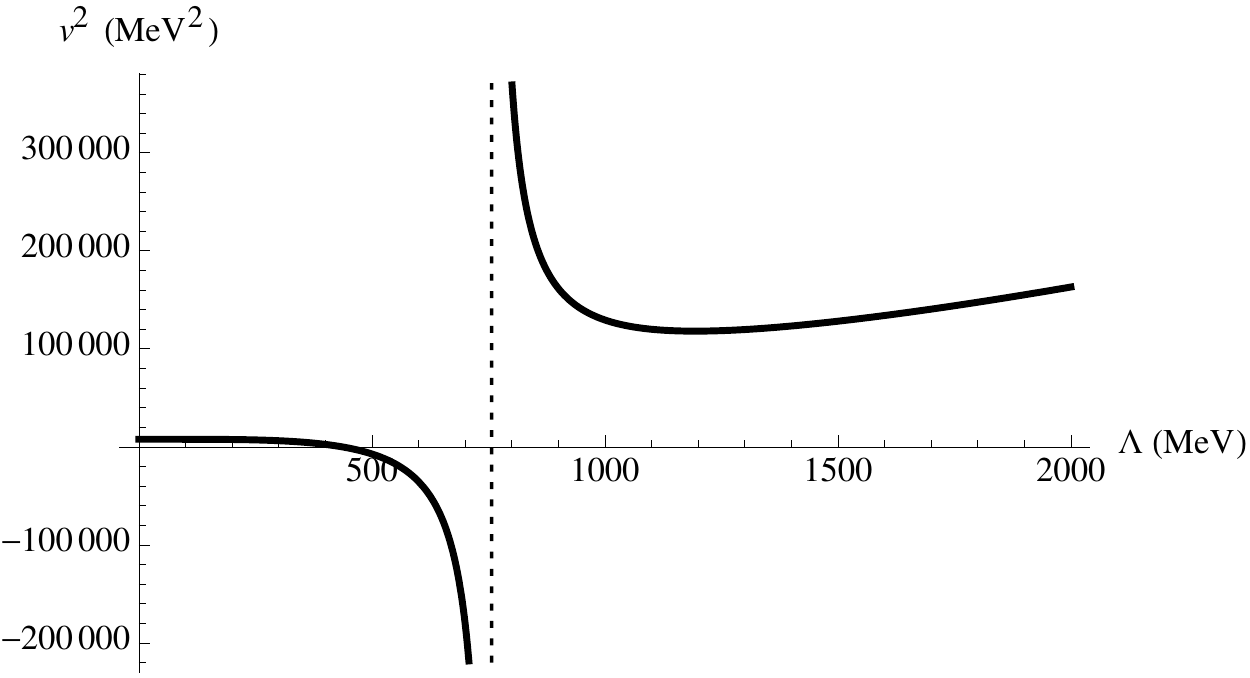}

\includegraphics[width=.3\textwidth]{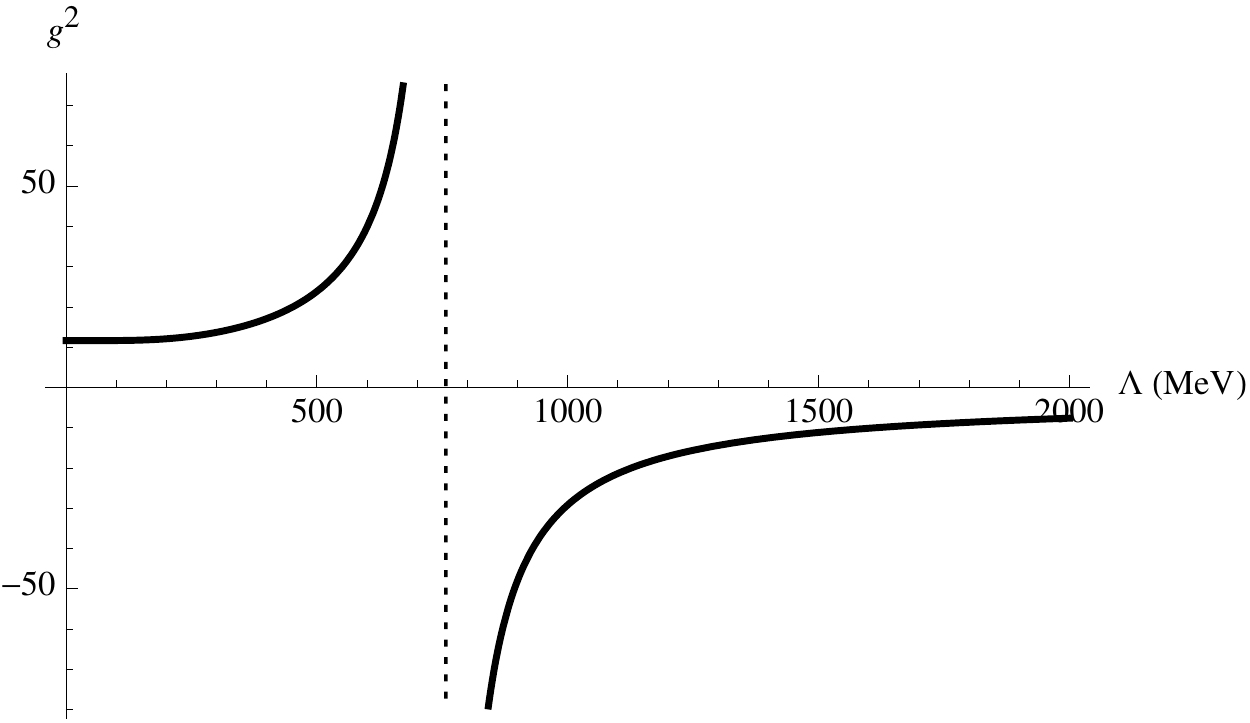}
\includegraphics[width=.3\textwidth]{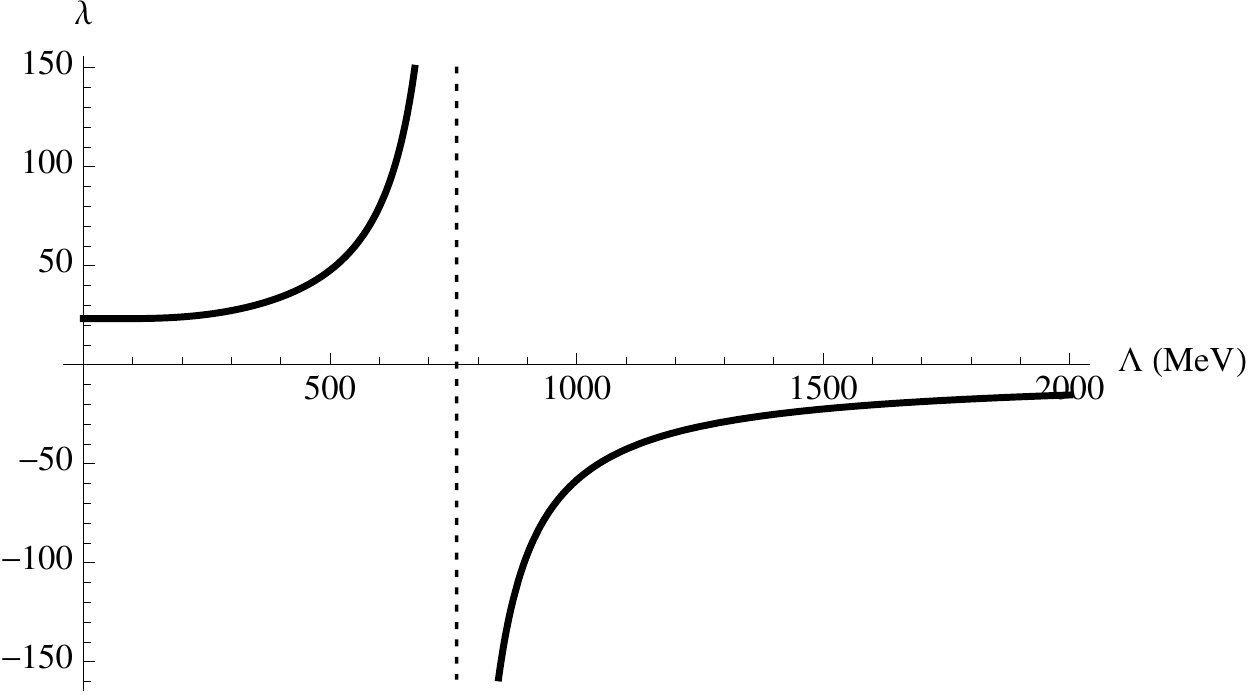}
\includegraphics[width=.3\textwidth]{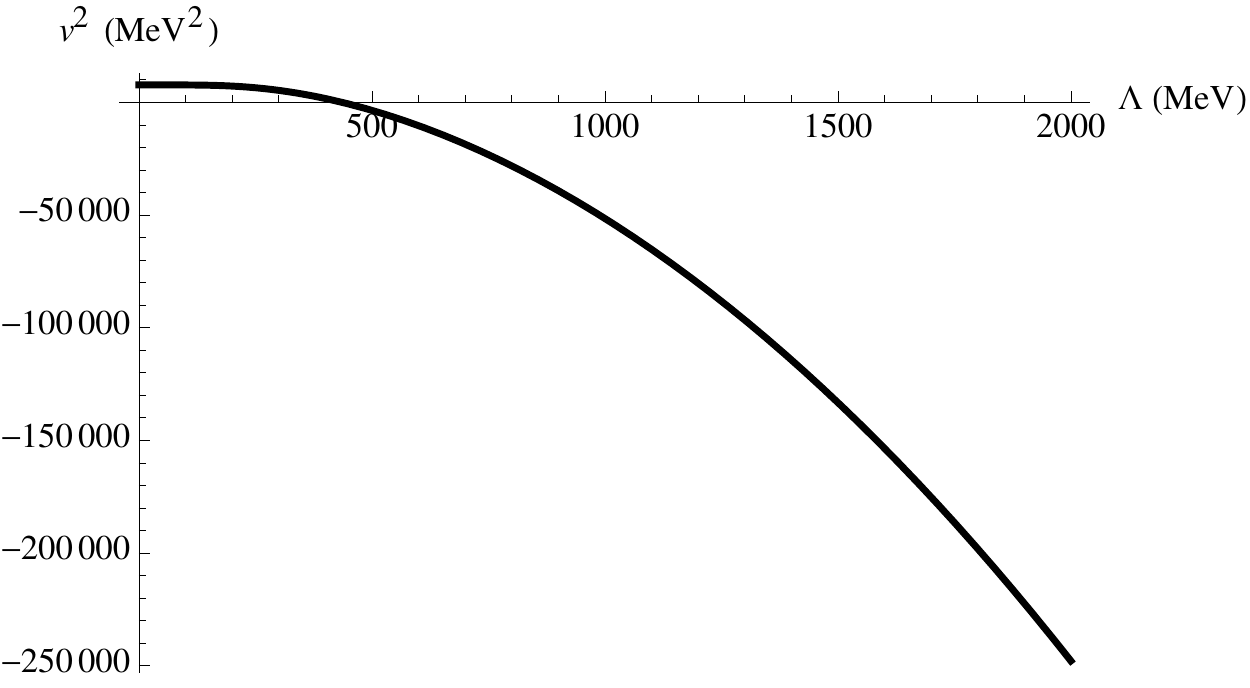}

\caption{Top row: $g^2$, $\lambda$ and  $v^2$ as functions of the PV cutoff parameter $\Lambda$
calculated within the BC scheme. Bottom: The same parameters in the RP scheme.
\label{fig:params} }
\end{figure}

By a quick comparison of  \Eqs{eq:g2BC} and (\ref{eq:lambdaBC}) with (\ref{eq:g2RP}) 
and (\ref{eq:lambdaRP}) one immediately sees that the behavior of the model parameters 
as a function of the regulator is dramatically different in the two prescriptions.  
For instance, $g^2$ is  $\Lambda$-independent in the BC scheme, while it strongly varies with it in the RP scheme, even exhibiting a pole structure \cite{Carignano:2014jla}. Similarly, 
the behaviors of $\lambda$ and $v^2$ are entirely different.
This is illustrated in \Fig{fig:params}, where, as in all our numerical examples, we have chosen 
the chiral limit value of the pion decay constant $f_\pi = 88$~MeV and a vacuum constituent quark mass of $\Mv=300$ MeV. 
For the sigma mass we took the input value $\ms = 600$~MeV in this figure.

In spite of these qualitative differences, the two prescriptions yield similar results for the phase diagram
if only spatially homogeneous mesonic fields are considered. 
In order to see this, 
we start from the thermodynamic potential $\Omega_\text{hom} (T,\mu; \Delta )$
and recall that only the meson contribution, \Eq{eq:UDelta}, depends explicitly on the model parameters. 
Hence, all differences between the thermodynamic potentials in the BC and RP prescriptions will stem from this term.

As a first step, using the gap equation we substitute the parameter $v^2$, which in both schemes is given by \Eq{eq:v2}. 
Inserting this into \Eq{eq:UDelta} we get
\beq
       U(\Delta) = \frac{\lambda}{4g^4} \left(\Delta^2 - \Mv^2\right)^2 + \frac{1}{2}\left(\Delta^2-\Mv^2\right) \Lonemv +
                         \frac{g^4}{4\lambda} \Lonemv^2 \,.
                         \label{eq:udeltacomp}
\eeq
From this, one can clearly see that the model parameters enter $U$ only
via the ratio $\lambda/g^4$, which will therefore be the focus of our discussion.
In the BC scheme we find (cf.~\Eqs{eq:g2BC} and (\ref{eq:lambdaBC}))
\beq
       \left(\frac{\lambda}{g^4}\right)_\text{BC} = \frac{m_\sigma^2 f_\pi^2 + 2\Mv^4 \Ltwomv}{2\Mv^4} \,,
\label{eq:lambdag4BC}
\eeq
while for the RP scheme this ratio can be expressed as 
\beq
     \left(\frac{\lambda}{g^4}\right)_\text{RP}  = 
         \left(\frac{\lambda}{g^4}\right)_\text{BC} + \eta(\ms^2)  \,,
\label{eq:lambdag4RP} 
\eeq
with
\beq
       \eta(\ms^2)  =  \left(\frac{m_\sigma^2}{4\Mv^2} -1\right)  \delta L_2(m_{\sigma}^2)  \,,
\label{eq:eta} 
\eeq 
where $\delta L_2(\ms) = L_2(\ms^2) - L_2(0)$ is a UV-finite quantity, see Appendix \ref{app:Lint}.
If $m_\sigma = 2\Mv$ it is immediate to see that 
$\eta$ vanishes and therefore the two expressions for $\lambda/g^4$
become identical.
This result is quite remarkable, as it means that in this case both schemes yield the same homogeneous phase diagrams,
despite the very different behavior of the model parameters.

\begin{figure}
\includegraphics[width=.32\textwidth]{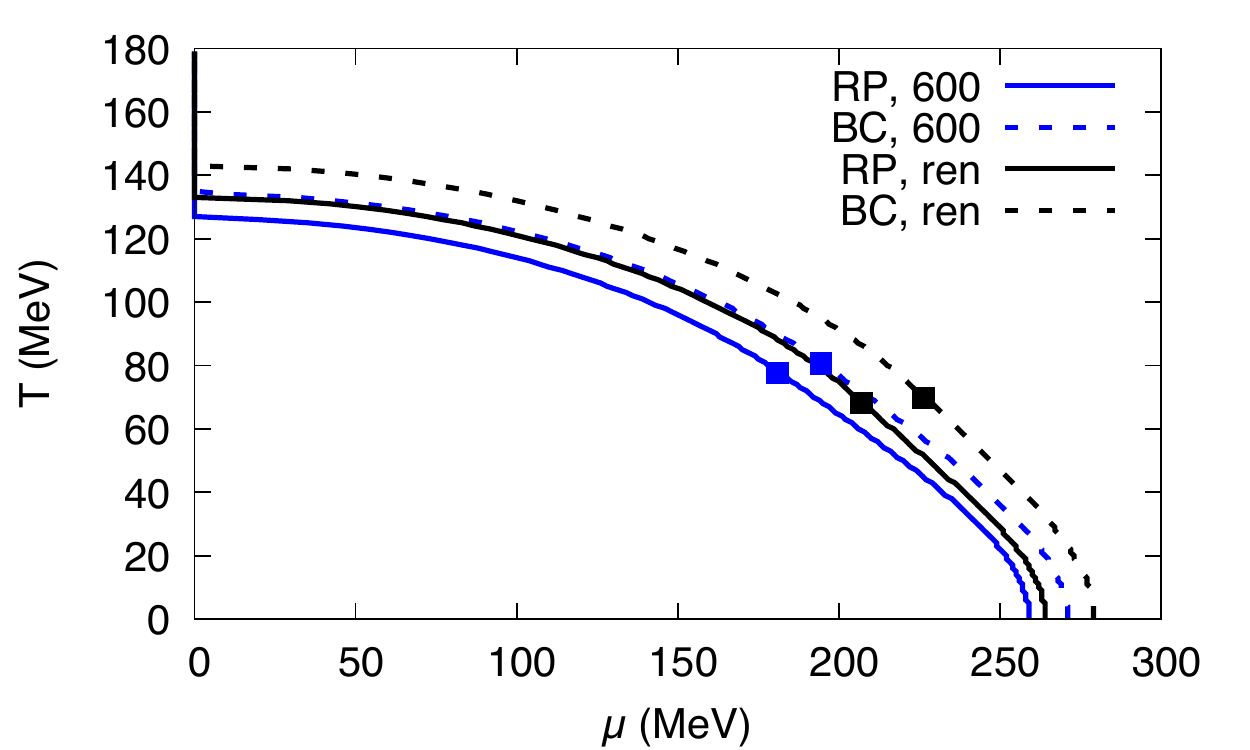}
\includegraphics[width=.32\textwidth]{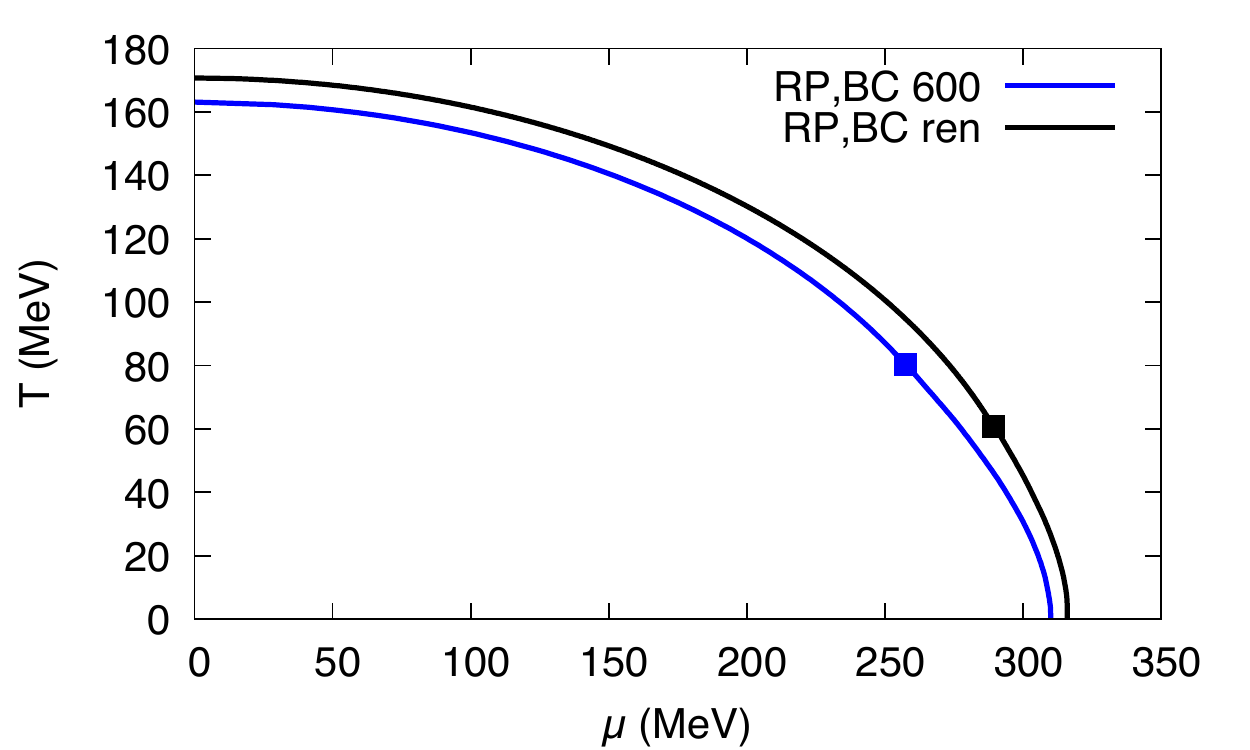}
\includegraphics[width=.32\textwidth]{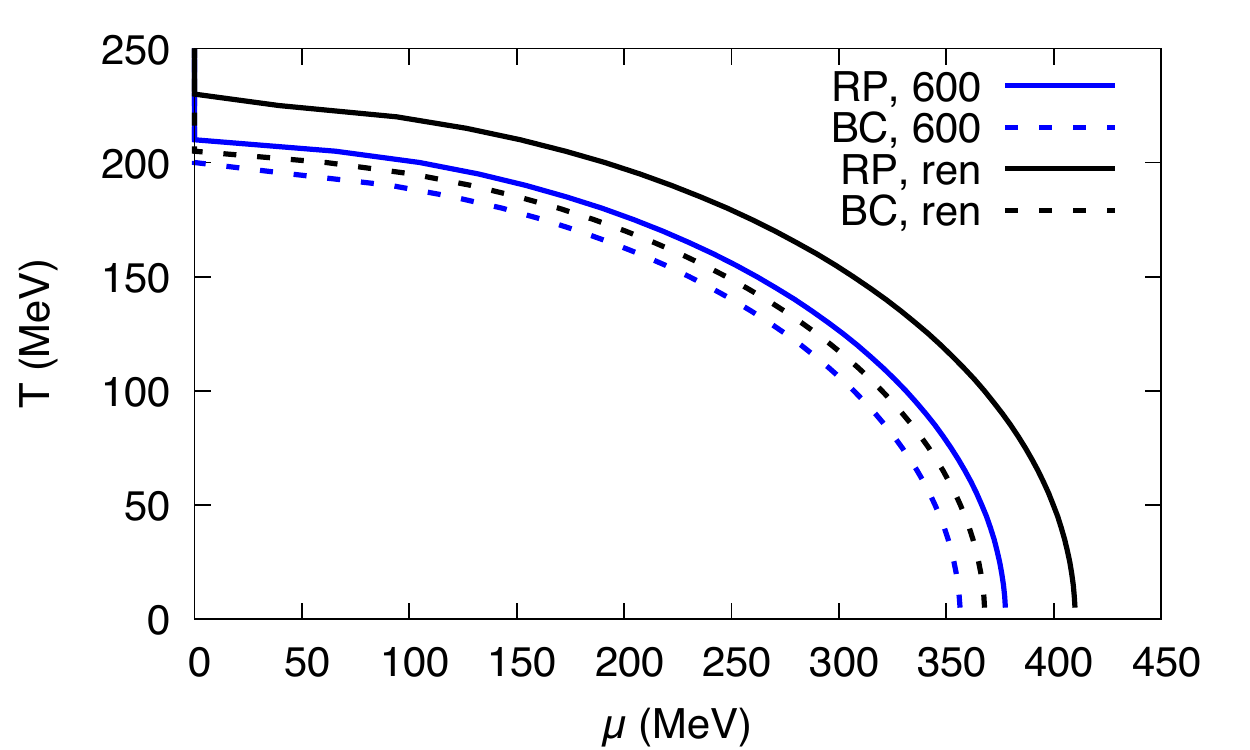}
\caption{Homogeneous phase diagrams for $\ms$ = 400 MeV (left), 600 MeV (center)  and 800 MeV (right) in the two schemes. Squares denote the CP. Note the different scales!  There is no CP for $\ms$=800 MeV.
\label{fig:pdms} }
\end{figure}

In the more general case $m_\sigma \neq 2\Mv$ on the other hand the two differ, so the resulting homogeneous phase diagrams will depend on the scheme of parameter fixing. 
In order to check this expectation, as well as determine the magnitude of this difference, we plot in \Fig{fig:pdms} the phase diagrams for three different values of $\ms$. 
While in the central panel, where $\ms = 2\Mv$, the curves obtained in the two schemes lie on top of each other, 
in the other cases, the two parameter fixing prescriptions lead to different results.

\begin{figure}
\includegraphics[width=.4\textwidth]{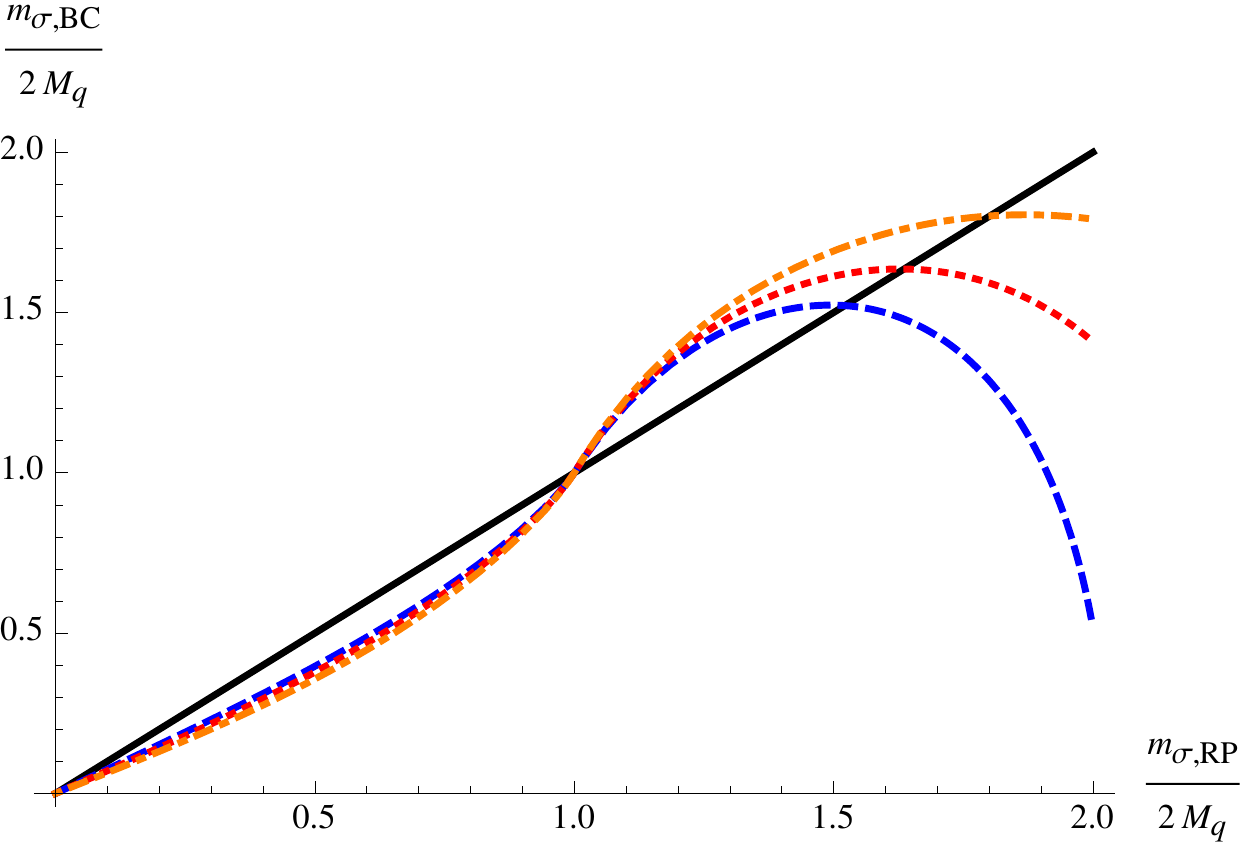}
\caption{Value of the sigma mass in the BC scheme which yields the same thermodynamic potential as the corresponding sigma mass in the RP scheme. The different curves correspond to different cutoff values: $\Lambda = 0$ (black solid line), $\Lambda = 700$ MeV (blue dashed line), $\Lambda = 1$ GeV (red dotted line) and the renormalized result (orange dash-dotted line). 
\label{fig:msBCofRP} }
\end{figure}

The differences are however relatively small. In fact, it is possible to compensate for the change of the parameter fixing scheme by a moderate adjustment of the input sigma mass.
To see this, we fix the value of $f_\pi$ and the cutoff $\Lambda$, then go back to \Eqs{eq:lambdag4BC} and (\ref{eq:lambdag4RP})
and ask for which input $\ms = \msBC$ in the BC scheme the ratio $\lambda/g^4$ (and hence the homogeneous phase 
diagram) becomes equal  to the one obtained with $\ms = \msRP$ in the RP scheme.
We arrive at the relation
\beq
      \msBC^2 
       = \msRP^2 - \frac{2\Mv^4}{f_\pi^2}  \, \eta(\msRP^2)
\label{eq:msBCRP}
\eeq
which is illustrated
in \Fig{fig:msBCofRP}
for different values of $\Lambda$ . 
The black solid line corresponds to the result in the sMFA, where $\Lambda = 0$ and the two schemes (and consequently the two input masses) are identical. 
For a non-zero cutoff instead, $\msBC$ is in general different from  $\msRP$, but there are still three points
 where 
the curves cross the black solid line, meaning that the two masses agree.
These crossing points correspond to the values of the sigma mass where $\eta(\ms^2)= 0$, cf.~\Eq{eq:eta}.
In particular, 
as we have discussed above, this is always the case at $\ms = 2\Mv$. 
In addition, $\delta L_2(\ms)$ vanishes at $\ms = 0$, as well as at some higher value of $\ms$, which depends on the cutoff.\footnote{
In the renormalized limit, $\Lambda \rightarrow \infty$, the value of $r = \frac{\msRP}{2\Mv}$ at the third crossing point is 
implicitly given by the equation
$r = \cosh(\frac{r}{\sqrt{r^2-1}})$, which yields $r \approx 1.81$.
}
As a consequence, $\msBC$ is always relatively close  $\msRP$, at least for reasonable values of the sigma mass.
Recalling that the latter is not well constrained by experiments, 
this means that for homogeneous matter the analysis in the theoretically inconsistent BC scheme 
incidentally leads to  results which are in practice indistinguishable from those of the correct treatment within the RP approach. 

The situation is however dramatically different for inhomogeneous phases, as we will now show for the CDW solutions.
Again, just like in the homogeneous case, only the  meson contribution $\Omega^{CDW}_\text{mes}$ to the thermodynamic
potential depends explicitly on the parameters, but the  essential difference here is the presence of the additional kinetic term 
in \Eq{eq:mesonpotCDW}. 
This term only contains the parameter $g$, which varies between different schemes, and unlike for the case of homogeneous matter no other parameter can compensate for it.
We thus conclude that inhomogeneous phases are sensitive to the way of parameter fixing, even in the case
$m_\sigma = 2\Mv$, where the homogeneous results are the same in both schemes.
In particular, while it was found that in the RP scheme the inhomogeneous phase survives in the renormalized limit \cite{Carignano:2014jla}, in the BC  scheme it quickly disappears as $\Lambda$ is increased.

\begin{figure}
\includegraphics[width=.3\textwidth]{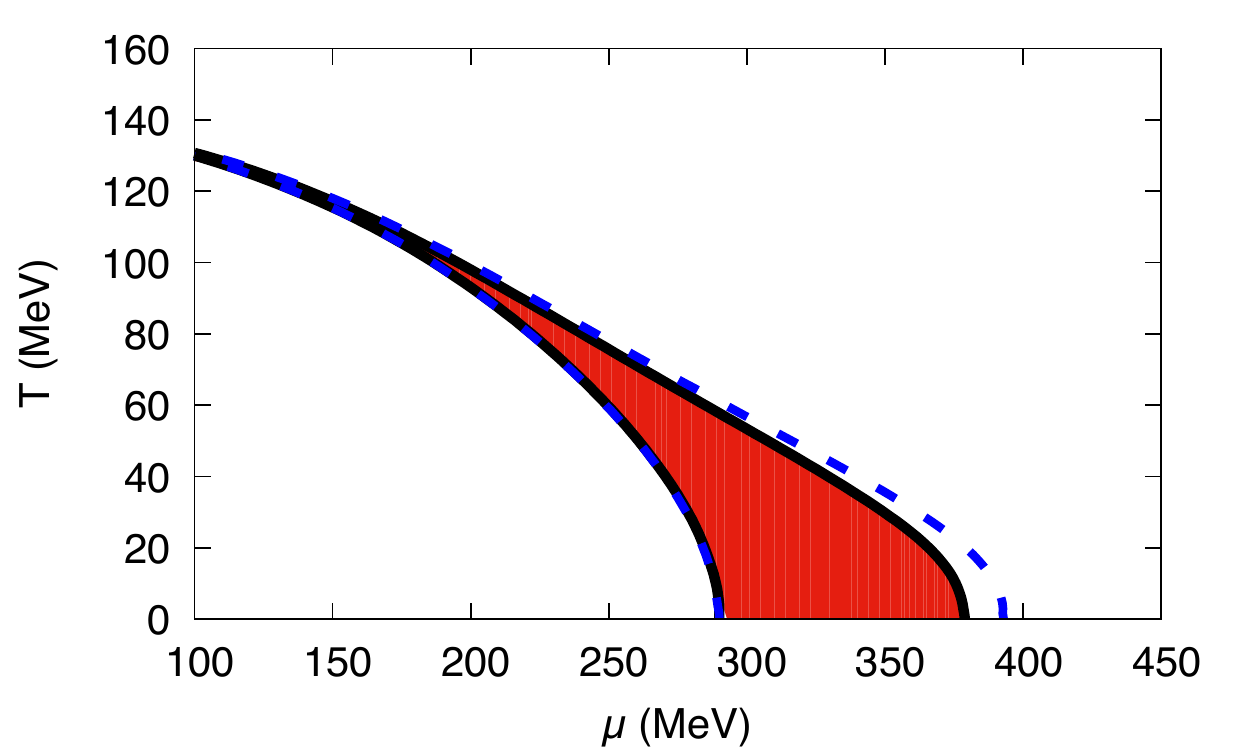}
\includegraphics[width=.3\textwidth]{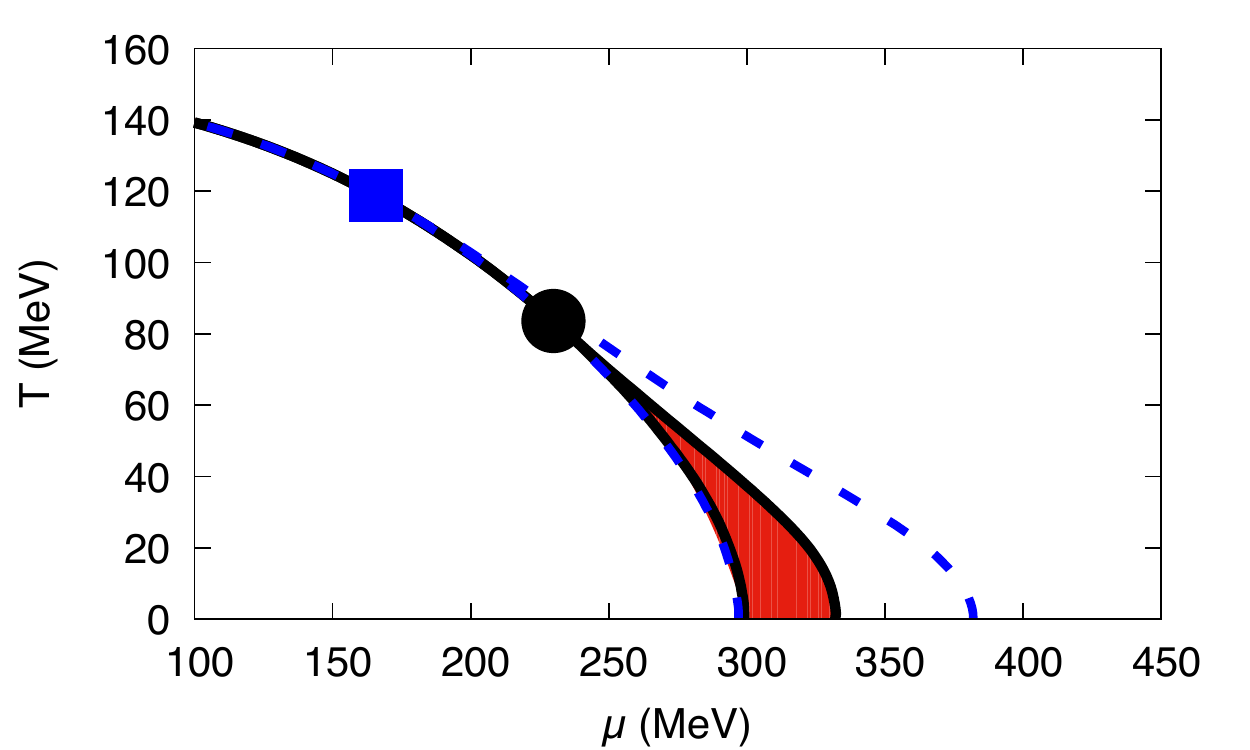}
\includegraphics[width=.3\textwidth]{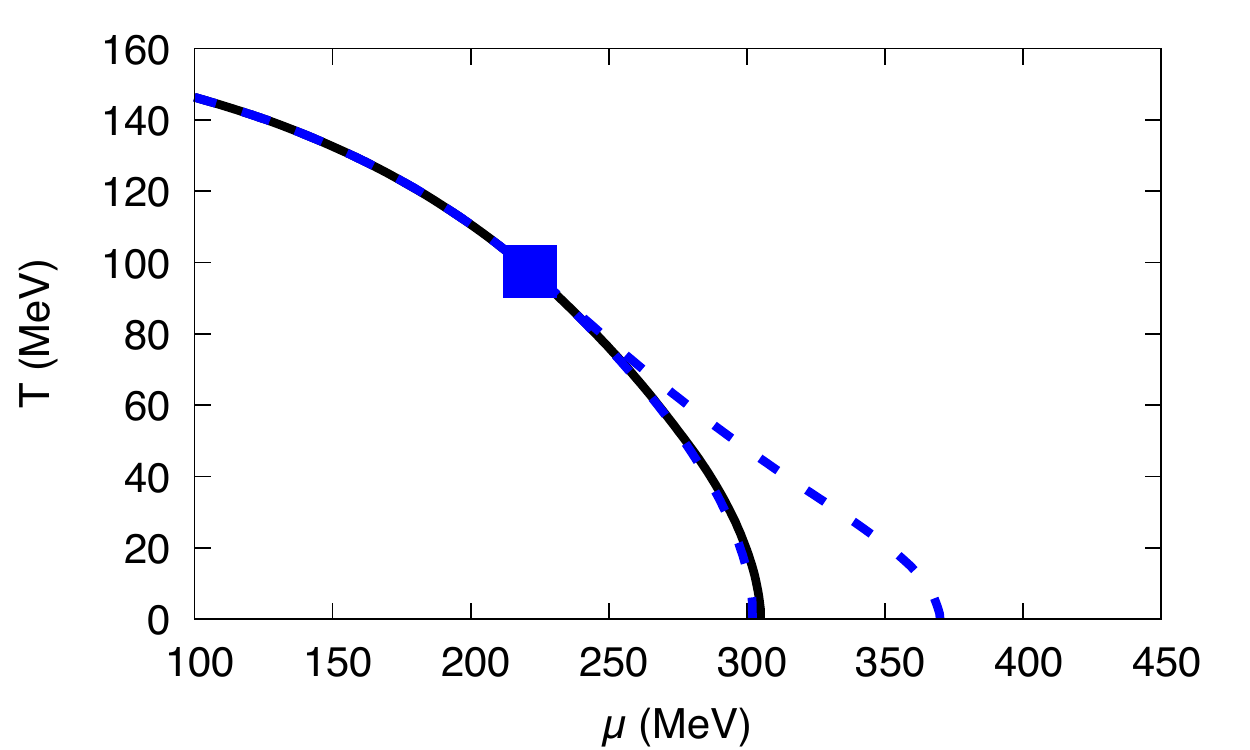}
\caption{Phase diagram for 
$\ms = 2\Mv$ and
$\Lambda=200$~MeV (left), 300~MeV (center) and 400~MeV (right). 
The shaded regions correspond to the inhomogeneous phases in the BC scheme, the blue dashed lines
indicate the boundaries of the inhomogeneous phase in the RP scheme.
Squares denote
the critical points, which 
are identical in both schemes and which
in the RP scheme coincide with the Lifshitz points. The dot indicates the Lifshitz point in the BC method.} 
\label{fig:pdinhom} 
\end{figure}

This can be seen in Fig.~\ref{fig:pdinhom} where the phase diagram for $\ms = 2\Mv$ is shown for three relatively small 
values of the cutoff.
When $\ms = 2\Mv$ the homogeneous phase diagrams are identical in both schemes, so that
the positions of the corresponding CPs (marked by the blue square) are the same.
Furthermore, in this limit it was found that within the RP scheme the LP coincides with the CP
\cite{Carignano:2014jla}. Consequently, the inhomogeneous phase 
  (whose boundaries
are indicated by the dashed lines) extends up to that point and is only mildly affected by the increase of the cutoff.
In the BC scheme, on the other hand, the LP (black circle) splits from the CP and moves towards the chemical-potential axis
when $\Lambda$ is increased. As a consequence, the inhomogeneous phase (shaded area) shrinks quickly, and already
at $\Lambda = 400$~MeV it has disappeared completely.
This behavior can be traced back to the fact that in the BC scheme, where the curvature mass $\msc$ is kept fixed,  
the pole mass $\msp$ decreases with increasing $\Lambda$. The observed shrinking of the inhomogeneous phase is then consistent with the results of Ref.~\cite{Carignano:2014jla},
where it was found that in the RP scheme the size of the inhomogeneous region decreases quickly when value of $\msp$ is decreased.

\section{Ginzburg-Landau analysis for the renormalized limit} 
\label{sec:renormGL}

It is worthwhile at this point to investigate the ``renormalized'' behavior of the model, i.e., the limit in which the regulator $\Lambda$ is sent to infinity, in the two different
prescriptions. Rather than presenting a numerical analysis of the full thermodynamic potential, we will focus on the GL coefficients,
for which simple analytical expressions can be obtained in the renormalized limit.  In particular, we will investigate the 
stability of the position of the CP and the LP as the cutoff increases, and the differences between the two schemes.

The relevant coefficients associated with 
 the position of the chiral critical point are $\gamma_2$ and $\gamma_{4,a}$, cf. \Sect{sec:qmmodel}.
Using again the gap equation \Eq{eq:v2}, we can rewrite the mesonic $\alpha_2$ contribution as 
\beq
       \alpha_2 = -\frac{\lambda}{g^4} \Mv^2  + \Lonemv\,,
       \label{eq:alpha2gapeq}
\eeq
and thus find, consistently with the discussion in the previous section for homogeneous phases,
that $\alpha_2$ and $\alpha_{4,a}$ (\Eq{eq:glalphas})
depend only on the ratio $\frac{\lambda}{g^4}$.
 It is then straightforward to see that the difference between the two parameter fixing prescriptions for both $\gamma_2$ and $\gamma_{4,a}$
is proportional to the difference $\eta$ between these ratios, see \Eqs{eq:lambdag4RP} and (\ref{eq:eta}),
which is a UV-finite quantity. More specifically, one  finds 
\begin{align}
\gamma_2^\text{RP}-\gamma_2^\text{BC}  & =     - \Mv^2 \,\eta(\ms^2) \,,
      \\
      \gamma_{4,a}^\text{RP}-\gamma_{4,a}^\text{BC}  & =   \eta(\ms^2) \,.
      \label{eq:diffGL}
  \end{align}

Having determined the differences between the coefficients for an arbitrary $\Lambda$, we now calculate their renormalized limit. 
 For this, we focus on the BC scheme and using
\Eqs{eq:glbetasvac}, (\ref{eq:lambdag4BC})
we obtain for $\gamma_2$ 
\begin{alignat}{1}
\gamma_2^\text{BC} &= - \frac{f_\pi^2 \ms^2}{2\Mv^2} -  \Ltwomv \Mv^2 + \Lonemv - \Lonez  + \beta_2^\text{med}
\nonumber \\
&\lraunder{\Lambda\rightarrow\infty}  - \frac{f_\pi^2 \ms^2}{2\Mv^2}  -\frac{ N_f N_c}{4 \pi ^2} \Mv^2  + \beta_2^\text{med} \,,
\label{eq:g2vac}
\end{alignat}
where the quadratic divergences cancel among the $L_1$ integrals, while the $L_2$ integral takes care of the remaining logarithmic terms, so that $\gamma_2$ is always UV-finite. 
Similarly, for the $\gamma_{4,a}$ coefficient one has
\begin{alignat}{1}
\gamma_{4,a}^{\text{BC}} 
&=
  \frac{f_\pi^2 \ms^2}{2\Mv^4} +   \Ltwomv  - \Ltwoz + \beta_4^\text{med}
\nonumber\\  
&\lraunder{\Lambda\rightarrow\infty}
\frac{f_\pi^2 \ms^2}{2\Mv^4} +  \frac{ N_f N_c}{4 \pi ^2} \log\frac{\Mv^2}{\epsilon^2}  + \beta_4^\text{med} \,,
\end{alignat}
where in the same way as before the logarithmic divergences in the UV cancel out between the $L_2$ integrals.
Since $\Ltwoz$ has an additional logarithmic divergence in the IR, we  intermediately
introduced a small regulator mass  
$\epsilon$.
This divergence is  in any case cancelled by the medium contribution $\beta_4^\text{med}$, 
so that after combining these two terms the limit $\epsilon \rightarrow 0$
can be taken numerically. 
From these results, the renormalized limit for the coefficients in the RP prescription can now straightforwardly be obtained using  \Eq{eq:diffGL}.

Once again, the situation changes drastically when considering inhomogeneous phases. There the relevant GL coefficient
$ \alpha_{4,b}$ (see \Eq{eq:glalphas}) carries a dependence on the parameter $g$ only, which has a completely different behavior in the two schemes.
Combining it with the fermionic part,
one has
\beq
\gamma_{4,b} = \frac{2}{g^2} - \Ltwoz + \beta_4^\text{med}\,,
\eeq
so that in order to obtain a UV-finite result, the first term on the right-hand side
must compensate the logarithmic divergence in $\Ltwoz$. This is the case for the RP scheme, where one 
obtains from \Eq{eq:g2RP}
\begin{alignat}{1}
\gamma_{4,b}^\text{RP} &= 2\frac{f_\pi^2}{\Mv^2} + \Ltwomv - \Ltwoz + \beta_4^\text{med}
\nonumber \\
&\lraunder{\Lambda\rightarrow\infty}
2\frac{f_\pi^2}{\Mv^2} 
+ \frac{ N_f N_c}{4 \pi ^2} \log\frac{\Mv^2}{\epsilon^2}  + \beta_4^\text{med} \,,
\end{alignat}
which is again UV-finite
(and for  $\ms = 2\Mv$
coincides with $\gamma_{4,a}^\text{RP}$~\cite{Carignano:2014jla}
and $\gamma_{4,a}^\text{BC}$). 

On the other hand, in the BC scheme $g^2$ is simply a constant, cf. \Eq{eq:g2BC}, so that 
 there is no additional term available to cancel the UV divergence in 
$\Ltwoz$, 
 and the $\gamma_{4,b}$ coefficient diverges logarithmically with $\Lambda$.
 Recalling that $\gamma_{4,b}$ is proportional to the coeeficient of the $q^2$ term in the GL expansion of the
 thermodynamic potential, cf.~\Eq{eq:OmegaGL},\footnote{
 Since $\Ltwomv$ is negative, gradient terms become strongly disfavored
 at large cutoff values, which is consistent with the disappearance of the inhomogeneous phase seen in 
 Fig.~\ref{fig:pdinhom}. 
} 
we thus conclude that a renormalized limit of $\Omega$ does not exist, and hence 
the BC scheme is completely inadequate for dealing with inhomogeneous phases.

\section{Other parameter fixing schemes}
\label{sec:otherschemes}

Until now, we focused our discussion on the two most commonly employed parameter fixing schemes discussed in the literature. 
Since the two differ in both the identification of the sigma mass and the pion decay constant, one might at this point 
think of introducing ``hybrid'' schemes which mix the prescriptions employed in the two schemes discussed.

A first possibility would be to identify $f_\pi$ with the vacuum sigma expectation value as in the BC scheme, 
but associate $\ms$ with its pole value.
The corresponding equations for the parameters  $g$ and $\lambda$ in this ``BP'' scheme 
are 
\beq
\label{eq:params_BC_g}
g^2 = \frac{\Mv^2}{f_\pi^2} 
\eeq
and
\begin{align}
\label{eq:params_BC_lambda}
\lambda &= g^4 \left[ \frac{\ms^2}{2 \Mv^2 g^2} +\left( 1 - \frac{\ms^2}{4\Mv^2} \right) \Ltwomsmv \right] 
\nonumber\\
&= \frac{\Mv^4}{f_\pi^4} \left[ \frac{\ms^2 f_\pi^2}{2 \Mv^4} +\left( 1 - \frac{\ms^2}{4\Mv^2} \right) \Ltwomv -     \eta(\ms^2)  \right]  \,,
\end{align}
from which $v$ can again be obtained via the gap equation, cf.~\Eq{eq:v2}.

By a quick inspection of these expressions we see immediately that for the case $\ms = 2\Mv$ one has $\lambda/g^4 = 2/g^2$, so that,
 according to \Eq{eq:glalphas}, $\alpha_{4,a}^\text{BP}$ agrees with  $\alpha_{4,b}^\text{BP}$ and thus
the CP coincides with the LP, just like in the RP scheme.\footnote{In fact, this is consistent with the finding of \cite{Carignano:2014jla} that for $\msp = 2\Mv$ the two points agree,
irrespective of the choice of $f_\pi$.}
The UV behavior of the GL coefficients is however  very different. Focusing for simplicity on  
$\gamma_2$, using \Eq{eq:glalphas} and comparing the result with \Eq{eq:g2vac}
we find
\beq
\gamma_2^\text{BP} = \gamma_2^\text{BC} + \Mv^2\, \eta(\ms^2) + \frac{\ms^2}{4}\Ltwomv\,,
\eeq
where the first two terms on the right-hand side are finite.
The last term however  diverges logarithmically when the cutoff is sent to infinity,
and therefore $\gamma_2^\text{BP}$ diverges as well.
According to \Eq{eq:OmegaGL} and since $\Ltwomv$ is negative, this means that $M=0$ corresponds to a maximum
of the thermodynamic potential, so that for $\Lambda \rightarrow \infty$ chiral symmetry never gets restored,
even at arbitrarily high temperatures or chemical potentials. 

The other possible hybrid scheme one can consider involves fixing the pion decay constant to its renormalized value as in the RP scheme, but identifying the sigma 
curvature mass
with its physical value. This ``RC'' scheme gives 
\beq
\label{eq:params_RC_g}
g^2 = \frac{\Mv^2}{f_\pi^2 + \frac{1}{2} \Mv^2 \Ltwomv} 
\eeq
and
\begin{align}
\label{eq:params_RC_lambda}
\lambda &= g^4 \left[ \frac{\ms^2}{2 \Mv^2 g^2} + \Ltwomv \right] 
\nonumber\\
&= \left( \frac{\Mv^2}{f_\pi^2 + \frac{1}{2} \Mv^2 \Ltwomv} \right)^2  \left[ \frac{\ms^2}{2\Mv^4} \left(f_\pi^2 + \frac{1}{2} \Mv^2 \Ltwomv \right) + \Ltwomv \right] \,.
\end{align}
For the GL coefficient $\gamma_2$ one obtains 
\beq
\gamma_2^\text{RC} = \gamma_2^\text{BC} - \frac{\ms^2}{4}\Ltwomv\,,
\eeq
which again has a logarithmic divergence in the same form as in the BP scheme, 
but with  opposite sign.
As a consequence, $M=0$ corresponds to a local minimum of the thermodynamic potential at large cutoff values.
At the same time,  at large $\Lambda$ the vacuum solution $M= \Mv$ of the gap equation corresponds to a maximum at  large 
$\Lambda$, so that beyond a critical cutoff value in this scheme chiral symmetry is never broken, even in vacuum.

We thus conclude that neither of these hybrid schemes is appropriate for the study of the QM model in the eMFA,
 not even when the analysis is restricted to homogeneous phases.

\section{Conclusions}
\label{sec:conclusions}

In this work we investigated the sensitivity of the chiral phase structure of the quark-meson model,
with a particular emphasis on inhomogeneous phases,
to the  parameter fixing prescription in an extended mean-field approximation
where fermionic vacuum fluctuations are taken into account.
In the most commonly employed prescription, which we  referred to as BC scheme,
the pion decay constant and the sigma-meson mass are identified with the vacuum expectation value of the sigma field and the 
curvature mass, respectively. 
While correct in the standard mean-field approximation,
where fermionic vacuum fluctuations are neglected, this prescription is however
inconsistent in the extended mean-field approximation.  This is due to the fact that the quark loops 
not only  change the ground-state of the model, but also give rise to a renormalization of the
meson masses and wave functions. 
The proper procedure, which we termed RP scheme, therefore  involves the 
renormalized pion decay constant and the sigma pole mass.

Although these two parameter fixing schemes lead to very different behaviors of the model parameters $g^2$, $\lambda$ and
$v^2$ as functions of the regulator, we found that this has only a relatively small effect on homogeneous phases:
In this case,   the thermodynamic potential depends only on the ratio $\lambda /g^4$, for which the differences between the individual
couplings in the two schemes cancel each other to a large extent.  In particular for $\ms = 2\Mv$, i.e., 
if the sigma mass is chosen to be equal to twice the constituent quark mass in vacuum, $\lambda /g^4$ is equal in both schemes,
leading to exactly the same homogeneous phase diagram. For  $\ms \neq 2\Mv$  differences exist  but are still relatively 
small, so that equal phase diagrams can be obtained by choosing slightly different values of $\ms$ in the two schemes.
As a consequence, since $\ms$ is not well constrained by experiments, 
the analysis of  homogeneous phases in the theoretically inconsistent BC scheme 
incidentally leads to  results which are in practice indistinguishable from those of the correct RP approach.

For inhomogeneous phases the situation is however completely different and the choice of the proper parameter fixing scheme becomes crucial. 
While in the RP scheme the inhomogeneous phase was found to be relatively stable when fermionic vacuum fluctuations are
included~\cite{Carignano:2014jla}, in the BC scheme it disappears already at rather low values of the regulator $\Lambda$. 
In particular,  the coincidence of the CP with the LP in the case $\ms = 2\Mv$ is only present when the pole mass is
employed in the parameter fixing.

We obtained further insights into this behavior by performing a Ginzburg-Landau analysis in the renormalized
limit, where the regulator $\Lambda$ is sent to infinity.
We found that the coefficients $\gamma_2$ and $\gamma_{4,a}$, which determine the location of the critical point in the homogeneous phase diagram, remain finite in both schemes. On the other hand, the coefficient $\gamma_{4,b}$, which is relevant for the Lifshitz point in the inhomogeneous case, turned out to be only finite in the RP scheme, whereas it diverges in the BC scheme. Since the GL coefficients are obtained from a Taylor expansion of the thermodynamic potential, this means that
in the BC scheme the model does not even have a well defined renormalized limit if inhomogeneous phases are considered.
Within the same  GL framework  we also studied two possible ``hybrid'' schemes, where either the bare pion decay constant and the sigma pole mass
or the renormalized $f_\pi$ and the curvature mass are used. 
We found that in these schemes even the coefficient $\gamma_2$ becomes UV divergent, meaning that the
renormalized limit does not even exist if the model is restricted to homogeneous mean fields.

We thus conclude that in the
presence of quark vacuum contributions,
 the parameters in the QM model should be fixed according to the RP scheme,
especially when dealing with inhomogeneous phases.
Although our analysis was restricted to the extended mean-field approximation, where only fermionic vacuum fluctuations
are taken into account, we believe that this result also applies to more advanced approximations, e.g., for calculations performed within the framework of the functional renormalization group.

Finally we recall that the sigma meson mass and in particular the constituent quark mass do not correspond to good
observables in reality. It might thus be worthwhile to think of better alternatives for future applications of the model.

\appendix

\section{Regularized loop functions}
\label{app:Lint}

The loop functions which often enter our calculations are 

\beq
\label{eq:L1def}
\Lonem(M)  = 4iN_f N_c \int\!\! \frac{d^4 p}{(2\pi)^4} \frac{1}{p^2 - M^2 +
  i\epsilon} = 2N_f N_c  \int\!\! \frac{d^3 p}{(2\pi)^3} \frac{1}{\sqrt{\p^2 + M^2}} \,,
\eeq 
and
\bea
\label{eq:L2def}
       L_2(q^2; M) 
       &=& 4iN_f N_c \int\!\! \frac{d^4 p}{(2\pi)^4} \frac{1}{[(p+q)^2
         - M^2 + i\epsilon][p^2 - M^2 + i\epsilon]} 
\nonumber\\       
       &=& 4N_f N_c \int\!\! \frac{d^3 p}{(2\pi)^3} \frac{1}{\sqrt{\p^2 + M^2}}
                \frac{1}{q^2 - 4(\p^2+M^2) + i\epsilon}\,,
\eea
which can be conveniently split into $L_2(q^2) = L_2(0) + \delta L_2(q^2)$, where only the last term can develop an imaginary part.
Since these quantities are related to vacuum quark loop integrals, the argument $M$ appearing in their expressions is typically the constituent quark mass in vacuum $\Mv$.  
In light of this, throughout the paper we will omit this argument by defining $L_1(\Mv) \equiv L_1$ and $L_2(q^2,\Mv) \equiv L_2(q^2)$, and will explicitly write when the argument is different
(eg. in the expressions for the GL coefficients, where $M = 0$).

Employing Pauli-Villars regularization with three regulators we get the following explicit expressions 
for the loop functions~\cite{Carignano:2014jla}: 

\begin{align}
\Lonem&\rightarrow
\frac{N_f N_c}{4\pi^2} \sum_{j=0}^3 c_j M^2_j \ln
M^2_j\,,
\\[1mm]
\Ltwom  &\rightarrow
\frac{N_f N_c}{4\pi^2} \sum_{j=0}^3 c_j \ln M^2 _j\,,
\\[1mm]
\mathrm{Re}\,\delta L_2(q^2)  &\rightarrow
\frac{N_f N_c}{4 \pi^2}
\sum_{j=0}^3 c_j \, \left\{ f\left(\frac{4 M^2_ j}{q^2} -1\right)
-2 \right\},
\\[1mm]
\mathrm{Im}\,\delta L_2(q^2) &\rightarrow
-\frac{N_f N_c}{4 \pi}
\sum_{j=0}^3 c_j \, \sqrt{1 -\frac{4 M^2_ j}{q^2}} \,
\theta(q^2-4M_j^2)\,,
\end{align}

with $M_j^2 = M^2 + j\Lambda^2$ and
\beq f(x) = \left\{ \begin{array}{ll}
    2\sqrt{x} \arctan(\frac{1}{\sqrt{x}})\,, & \quad x > 0
    \\[1mm]
    \sqrt{-x} \ln(\frac{1+\sqrt{-x}}{1-\sqrt{-x}})\,, & \quad x < 0
    \\[1mm]
    0\,, & \quad x=0\,.
  \end{array} \right.
\eeq

For $\Lambda \to \infty$, it is easy to see that $\Lonem$ diverges like $\sim \Lambda^2 + M^2\log(\Lambda)$, while $\Ltwom$ has only an $M$-independent logarithmic $\sim \log(\Lambda)$ divergence. More specifically, one finds
\beq
\Lonem = \frac{N_f N_c}{4\pi^2} \left[ 3  \Lambda^2 \log\left(\frac{4}{3}\right) + M^2\left( \log\left(\frac{8 M^2}{3\Lambda^2}\right) - 1 \right) \right] + {\cal{O}}\left(\frac{1}{\Lambda^2}\right)
\eeq
and
\beq
\Ltwom = \frac{N_f N_c}{4\pi^2} \log\left(\frac{8 M^2}{3\Lambda^2}\right)+ {\cal{O}}\left(\frac{1}{\Lambda^2}\right)
\eeq

The $\delta L_2$ terms are instead UV-finite. 
In particular, for $\Lambda \to \infty$ one can see that all the regulator-dependent terms in $\mathrm{Re}\,\delta L_2(q^2)$ drop,  
effectively reducing it to its unregularized version:
\beq
\mathrm{Re}\,\delta L_2(q^2)  \lraunder{\Lambda\to\infty}
\frac{N_f N_c}{4 \pi^2}
\left\{ f\left(\frac{4 \Mv^2}{q^2} -1\right)
-2 \right\} \,.
\eeq

\end{document}